\documentclass[sigconf]{acmart}

\copyrightyear{2025}
\acmYear{2025}
\setcopyright{rightsretained}
\acmConference[RecSys '25]{Proceedings of the Nineteenth ACM Conference on Recommender Systems}{September 22--26, 2025}{Prague, Czech Republic}
\acmBooktitle{Proceedings of the Nineteenth ACM Conference on Recommender Systems (RecSys '25), September 22--26, 2025, Prague, Czech Republic}\acmDOI{10.1145/3705328.3759330}
\acmISBN{979-8-4007-1364-4/2025/09}

\usepackage{subcaption}
\usepackage{multirow}
\usepackage{xcolor}
\usepackage{framed}
\definecolor{bodybg}{RGB}{242,242,242}

\usepackage{tikz}
\usetikzlibrary{shapes}

\definecolor{bodybg}{RGB}{242,242,242}

\usepackage{float}
\usepackage{tikz}
\usetikzlibrary{calc}

\begin{document}

\title{Mitigating Popularity Bias in Counterfactual Explanations using Large Language Models}

\author{Arjan Hasami}
\orcid{0009-0005-6581-6013}
\affiliation{%
  \institution{Delft University of Technology}
  \city{Delft}
  \country{The Netherlands}
}
\email{a.hasami@student.tudelft.nl}

\author{Masoud Mansoury}
\orcid{0000-0002-9938-0212}
\affiliation{%
  \institution{Delft University of Technology}
  \city{Delft}
  \country{The Netherlands}
}
\email{m.mansoury@tudelft.nl}

\begin{abstract}
Counterfactual explanations (CFEs) offer a tangible and actionable way to explain recommendations by showing users a "what-if" scenario that demonstrates how small changes in their history would alter the system’s output. However, existing CFE methods are susceptible to bias, generating explanations that might misalign with the user's actual preferences. In this paper, we propose a pre-processing step that leverages large language models to filter out-of-character history items before generating an explanation. In experiments on two public datasets, we focus on popularity bias and apply our approach to ACCENT, a neural CFE framework. We find that it creates counterfactuals that are more closely aligned with each user's popularity preferences than ACCENT alone.
\end{abstract}



\maketitle

\definecolor{nonpop}{RGB}{93,165,218}
\definecolor{popular}{RGB}{225,225,225}
\def\barW{6}   
\def\h{0.6}    

\newcommand*\dualbar[3]{
  \pgfmathsetmacro{\wA}{#2*\barW}  \pgfmathsetmacro{\wB}{\barW-\wA}
  \begin{scope}[yshift=#1]
    \fill[nonpop]  (0,0) rectangle (\wA,\h);
    \fill[popular] (\wA,0) rectangle (\barW,\h);
    \draw          (0,0) rectangle (\barW,\h);
    \pgfmathtruncatemacro{\aPct}{round(#2*100)}
    \pgfmathtruncatemacro{\bPct}{100-\aPct}
    \node[anchor=east,font=\small] at (-0.2,\h/2) {#3};
    \node[font=\small] at (\wA/2,\h/2)          {\aPct\% non-popular};
    \node[font=\small] at (\wA+\wB/2,\h/2)      {\bPct\% popular};
  \end{scope}
}
\newcommand*\singlebar[2]{
  \begin{scope}[yshift=#1]
    \fill[popular] (0,0) rectangle (\barW,\h);
    \draw          (0,0) rectangle (\barW,\h);
    \node[anchor=east,font=\small] at (-0.2,\h/2) {#2};
    \node[font=\small] at (\barW/2,\h/2) {100\% popular};
  \end{scope}
}

\newcommand{\scatterplotcomp}[1]{%
\begin{tikzpicture}
\begin{axis}[
    xlabel={X},
    ylabel={Y},
    legend style={at={(0.5,-0.15)}, anchor=north, legend columns=-1},
    width=4cm,
    height=4cm,
    ymode=log,
    log basis y={10},
    mark size=2pt,
    grid=both,
]

\pgfplotstableread[col sep=comma]{#1}\loadedtable
\pgfplotstablegetrowsof{\loadedtable}
\pgfmathtruncatemacro{\nrows}{\pgfplotsretval-1}
\foreach \i in {0,...,\nrows} {
    \pgfplotstablegetelem{\i}{novel_x}\of{\loadedtable} \edef\nx{\pgfplotsretval}
    \pgfplotstablegetelem{\i}{novel_y}\of{\loadedtable} \edef\ny{\pgfplotsretval}
    \pgfplotstablegetelem{\i}{accent_x}\of{\loadedtable} \edef\ax{\pgfplotsretval}
    \pgfplotstablegetelem{\i}{accent_y}\of{\loadedtable} \edef\ay{\pgfplotsretval}
    \addplot [gray, thick, mark=none] coordinates {(\nx,\ny) (\ax,\ay)};
}

\addplot[
    only marks,
    color=blue,
    mark=*,
] table [
    x=novel_x,
    y=novel_y,
    col sep=comma,
] {#1};
\addlegendentry{Novel}

\addplot[
    only marks,
    color=red,
    mark=*,
] table [
    x=accent_x,
    y=accent_y,
    col sep=comma,
] {#1};
\addlegendentry{ACCENT}

\end{axis}
\end{tikzpicture}
}
\section{Introduction}
\label{section:introduction}
Recommender systems (RS) have become vital for guiding users to relevant content across various domains~\cite{recsys_multimedia, recsys_ecommerce}. As these systems become increasingly complex, leveraging deep learning and vast amounts of data, the decision-making process becomes more opaque, raising concerns about transparency and trust~\cite{trust_deep_learning}. This has resulted in the emergence of Explainable RS as an area of research. Various explanation approaches have been explored, but ensuring that explanations are faithful to the model's logic and meaningful to the user remains challenging~\cite{expl_machine_learning}. 

Counterfactual explanations (CFEs)~\cite{cf_exs, xrec_survey} have emerged as an appealing solution to the tangibility problem in explainable RS. A CFE identifies a small set of the user's past interactions such that if those interactions were removed, the system's top recommendation would change. These personalized and actionable explanations allow users to see how their tastes drive the outcomes. Indeed, user studies have found that people appreciate detailed CFEs, favoring them over generic transparency statements \cite{why_am_i_not_seeing_it}.

A well-known issue with CFEs is the Rashomon effect~\cite{interpretable_machine_learning}: multiple distinct explanations can be valid for the same recommendation, each highlighting a different set of changes that might contradict each other. While this diversity of possible explanations can be seen as a strength, it also raises the question of which explanation should be shown to the user. Prior work tackles this in several, non-mutually exclusive ways, such as keeping only the nearest edit~\cite{prince, countER, accent} or returning a diverse set of multiple counterfactuals~\cite{DiCE, grease}. Instead, we argue that the selected CFE should align with the user's expectations and interests. Highlighting an obscure or marginally relevant item may satisfy formal criteria, yet appear unintuitive or misleading to the user. Previous research also stresses that explanations should be tailored to the user and context, as different users value different criteria~\cite{effective_explanations,effect_transparancy_on_trust_art}. 

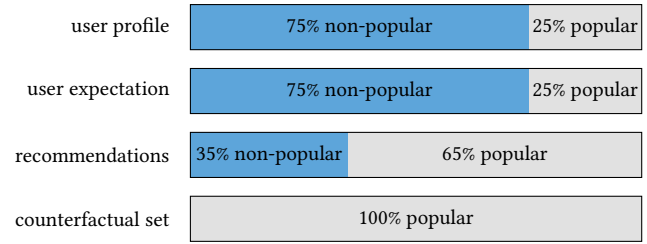
\begin{figure}[!t]
    \centering
    {\scriptsize
    \begin{tikzpicture}[x=1cm]
        \def\sep{0.25}
        \pgfmathsetmacro{\dy}{\h+\sep}
        \dualbar{0cm}{0.75}{user profile}
        \dualbar{-\dy cm}{0.75}{user expectation}
        \dualbar{-2*\dy cm}{0.35}{recommendations}
        \singlebar{-3*\dy cm}{counterfactual set}
    \end{tikzpicture}
    }
    \caption{A user with both popular and non-popular items in their history may expect a similar balance in recommendations and counterfactual explanations. However, this alignment often breaks due to popularity bias. The effect can be even stronger in counterfactual sets, which are typically small and thus more sensitive to such skew.}
    \Description{Four bars representing the distribution of popular and unpopular items in a user's profile (75\% non popular, 25\% popular), their expectation (75\% non popular, 25\% popular), recommendations (35\% non popular, 65\% popular) and a counterfactual explanation (100\% popular).}
    \label{figure:pop-expectation}
\end{figure}

The need for alignment becomes even more relevant when considering that explanation methods may reflect biases in the underlying recommender model. Because many CFE frameworks operate on internal signals such as gradients or influence measures, systemic biases can surface in the generated explanations. Figure \ref{figure:pop-expectation} demonstrates how this might look with popularity bias~\cite{unfairness_of_pop_bias,mansoury2020feedback,klimashevskaia2024survey,mansoury2022understanding}, where users mainly interested in niche items may fail to resonate with explanations based on popular items.

In this paper, we explore large language models (LLMs) as a tool to improve the user alignment of CFEs. LLMs possess a vast capacity to understand context, semantics, and nuances from textual data. Recently, LLMs have been leveraged in the recommendation task to turn a list of consumed items supplemented with descriptions into a coherent textual profile that captures themes a user consistently enjoys \cite{temporal_user_profiling}. We embed that profile, greedily remove history items whose absence changes it the most, and run ACCENT, a CFE framework for neural recommenders \cite{accent}, on this filtered history. 

By eliminating out-of-character interactions before the explanation step, we cut down the space of admissible CFEs and guide ACCENT toward explanations more closely resembling users' preferences. As this augmentation relies on a general "\textit{does this suit the user?}" test rather than on attribute-specific rules, it should, in principle, alleviate many kinds of mismatch. In this paper, we evaluate the approach solely on one well-known dimension, popularity bias. We present preliminary findings and show that it reduces mismatches between user popularity preferences and the popularity of items appearing in CFEs. Our contributions are threefold:

\begin{itemize}
    \item We propose two metrics for detecting popularity bias in counterfactual sets and use them to empirically confirm that ACCENT's explanations can be skewed.
    \item We propose LLM-augmented counterfactual generation, a plug-in pre-processing step that can be paired with any history-based CFE method.
    \item We show that explanations generated on the filtered history more closely resemble the user's popularity preference on two public datasets.
\end{itemize}
\section{Background}
\label{section:background}
\paragraph{Counterfactual Explanations in RS} Counterfactual explanations in recommender systems take the following form:

\begin{quote}
    You were recommended 'John Wick' because you previously liked 'Taken' and 'The Equalizer'. Otherwise, you would have been recommended 'Mean Girls'.
\end{quote}

By identifying a change in a user's history that flips the system's top-1 recommendation, CFEs aim to provide personalized, actionable feedback. Prior CFE frameworks have taken various approaches, such as optimization-based frameworks that solve a constrained minimization problem~\cite{countER, cf-gnn-explainer} or graph-centric approaches~\cite{prince, GNNUERS}. We adopt ACCENT~\cite{accent}, a framework based on influence analysis, as our primary method for generating CFEs.

\paragraph{ACCENT in a nutshell} ACCENT~\cite{accent} adapts influence functions~\cite{bbpreds,fastinfanal}--a technique from robust statistics~\cite{influence_curve}--to  produce CFEs for neural recommenders. For a given user $u$, let $r$ be the system's current top recommendation and $r^*$ an alternative item (taken from the original top-$k$ recommendations). ACCENT seeks the smallest subset $E$ of the user's interaction history $H_u$ whose removal would make the model prefer $r^*$ over $r$. Formally,
\begin{equation}
\text{ACCENT}(H_u,\hat{y}, r, r^*) = \operatorname*{arg\,min}_{E \subseteq H_u} \{|E|\bigm | \hat{y}_{u,r}^{-E} - \hat{y}_{u,r^*}^{-E} < 0\}
\end{equation}
\noindent where $\hat{y}^{-E}_{u,r}$ is the predicted score for user $u$ and item $r$ after virtually removing the interactions in counterfactual set $E$. ACCENT uses Influence functions to approximate these scores efficiently~\cite{fastinfanal}, avoiding retraining during candidate and subset evaluation.
\section{Problem Formulation}
\paragraph{Popularity bias in ACCENT} In recommender data, popular items converge the fastest~\cite{prent2024correcting}. As there are many more interactions, there are more training iterations and updates, giving them disproportionate weight during learning. ACCENT is based on influence functions, which tell us which training points are most used for a recommendation. However, due to the disproportionate weight of popular items due to popularity bias, these are more influential in most predictions. These popular items thus usually have a high influence score, making it likely that ACCENT picks them.

\paragraph{Towards bias-aware CFEs with LLMs} LLMs can encode high-level preferences, such as tone, themes, or narrative style, by reading free-text descriptions of consumed items. Recent work shows that LLM-based recommenders exhibit lower popularity bias than traditional baselines, and that careful prompting can reduce bias further \cite{llms_pop_bias}. We therefore investigate whether LLMs can act as a pre-processing step, filtering a user's history so that existing CFE frameworks such as ACCENT can start from a less biased candidate pool, leading to an explanation that aligns more with user interest.
\section{LLM-Augmented ACCENT}
User histories mix long-term tastes with one-off, context-driven interactions (e.g., watching a blockbuster with friends). These out-of-character items can skew CFE frameworks, such as ACCENT’s influence scores. To reduce this step, we first distill the user's core profile by removing inconsistent items, shrinking the search space, and dampening the popularity signal before the CFE step.

We leverage LLMs to generate natural language profiles of user preferences based on their full interaction history and item descriptions. The prompt asks for a concise (<300 words) summary focused on recurring themes, tones, or character types—avoiding broad genres or direct item mentions. We then embed the resulting text into a high-dimensional vector. An example prompt and LLM response is as follows:

\vspace{10pt}

\noindent
\begin{tikzpicture}
  \def\w{8.5cm}       
  \def\h{6cm}        
  \def\headerh{0.6cm} 
  \def\radius{6pt}    
  \def\hpad{15pt}     
  \def\vpad{2pt}      

  \fill[bodybg, rounded corners=\radius] (0,0) rectangle (\w,-\h);

  \begin{scope}
    \clip[rounded corners=\radius] (0,0) rectangle (\w,-\h);
    \fill[black] (0,0) rectangle (\w,-\headerh);
  \end{scope}

  \draw[line width=1.2pt, rounded corners=\radius] (0,0) rectangle (\w,-\h);

  \node[anchor=north west, text width=\w-2*\hpad, align=left, font=\fontsize{8.2}{10}\selectfont\bfseries\color{white}]
    at (\hpad,-\vpad)
    {Example LLM Query and Response (Movie Domain)};

  \node[anchor=north west, text width=\w-2*\hpad, align=justify, font=\fontsize{8.8}{10}\selectfont]
    at (\hpad,-\headerh-\vpad)
    {
    
    \textbf{Query:} \textit{Your task is to analyze a list of movies a user has interacted with and describe the profile and the preferences
    of the user in less than 300 words. Try to not be too broad
    (e.g. mention too many general categories such as action or
    comedy). Do not mention specific movie titles.}

    \vspace{0.7em}

    \textit{The user has watched the following movies: \\Die Hard (1988) - Action, Thriller - NYPD cop John...}

    \textit{...}

    \vspace{0.7em}


    \vspace{0.7em}

    \textbf{Response:} \textit{The user enjoys action-packed, high-stakes narratives with strong protagonists overcoming formidable odds.
    They prefer films that blend suspense, tension, and personal
    growth. There’s a clear affinity for classic cinema...}
    };
    \draw[dashed, black, line width=0.1pt] (0, -4.1) -- (\w, -4.1);
\end{tikzpicture}

\vspace{10pt}

Our goal is to find and remove items that, when excluded, significantly change this profile. The hypothesis is that filtering out inconsistent history items will help ACCENT produce CFEs more aligned with true user preferences.
First, we generate a textual profile $t_u$ of the user's preferences using an LLM prompted on their full interaction history and item descriptions: 
\begin{equation}
    \begin{alignedat}{2}
        t_u = \text{LLM}(\operatorname*{\text{concat}}_{i \in H_u}(p_i))
    \end{alignedat}
\end{equation}
\noindent then we embed $t_u$ into a high-dimensional vector $e_u$ using Sentence Transformer (SBERT)~\cite{reimers2019sentence}:
\begin{equation}
    \begin{alignedat}{2}
        e_u = \text{SBERT}(t_u)
    \end{alignedat}
\end{equation}
\noindent where $p_i$ is description of item $i$. We start with the full interaction history $H_u^{(0)} = H_u$. Then, for a fixed number of steps $n$ we iterate:

\begin{itemize}
    \item For every remaining item $i' \in H_u^{(j-1)}$ (history before step $j$), compute the embedding obtained after its removal $e_{u \setminus \{i'\}}^{(j-1)}$.
    \item Select the item whose removal produces the largest cosine dissimilarity with the current profile:
    \begin{equation}
        i_j^* = \operatorname*{\arg \max}_{i' \in H_u^{(j-1)}}(1-\cos(e_u^{(j-1)},e^{(j-1)}_{u \setminus \{i'\}}))
    \end{equation}
    \item Permanently discard $i_j^*$: $H_u^{(j)} = H_u^{(j-1)} \setminus \{i_j^*\}$.
\end{itemize}

After $n$ iterations, the set of all discarded items is $\tau_u = \{i_1^*,\dots,i_n^*\}$. We can finally create the filtered history $H_u^{(n)} = H_u \setminus \tau_u$ and feed it to ACCENT (or any other history-based CFE framework), ensuring explanations exclude these misaligned items. 
\section{Experiments}

This section outlines our experimental setup, including datasets, evaluation metrics, and baselines.

\label{section:experiments-and-results}

\begin{table}[t!]
    \centering
    \caption{Statistics of the datasets.}
    \vspace{-10pt}
    \begin{tabular}{lrrrr}
    \toprule
        \textbf{Datasets} & \textbf{\#users} &  \textbf{\#items} & \textbf{\#interactions} & \textbf{Density}\\
    \midrule
        ML-1M & 6,040 & 3,952 & 1,000,209 & 4.19\%\\
        Amazon & 94,761 & 25,612 & 814,587 & 0.03\%\\
    \bottomrule
    \label{tab:dataset}
    \end{tabular}
\end{table}



\subsection{Datasets} 
We evaluate our approach using two datasets: MovieLens 1M (ML-1M) \cite{movielens} and Amazon Video Games \cite{hou2024bridging}. Both datasets provide user-item interactions and rich item metadata, including textual descriptions and categories. This metadata is used to create the LLM prompts as previously discussed. For the Amazon dataset, we ensure all users and items have at least five interactions. Table \ref{tab:dataset} shows statistical properties of these datasets.

To keep LLM costs manageable, we focus on users whose histories are most vulnerable to popularity bias. Following previous research~\cite{unfairness_of_pop_bias}, we rank all users on the fraction of popular items in their history. We retain only the extremes: the top 20\% (\textit{blockbuster} users) and bottom 20\% (\textit{niche} users), discarding the middle 60\%. Users with more than 100 interactions are also dropped to curb runtime. From the survivors, we take the 250 most-niche and 250 most-blockbuster users per dataset to evaluate our approach\footnote{Filtering steps were only applied for generating counterfactuals; the recommendation models were trained on the full datasets.}.

\subsection{Baselines}

To evaluate LLM-augmented ACCENT, we compare its counterfactual sets against two baselines:
\begin{itemize}
    \item \textit{ACCENT}. We use the original ACCENT method \cite{accent} as the primary baseline, as our framework builds directly upon it. We use $k=5$, checking each of the top-5 recommendations for feasibility as a replacement for the top-1.
    \item \textit{Top Popular}. This heuristic baseline constructs counterfactual sets from the most popular items in a user's history, matching ACCENT's set size per user. The expectation is that this performs poorly for non-blockbuster users.
\end{itemize}

\subsection{Evaluation metrics}
\label{subsection:evaluation}
We report two complementary metrics to quantify how closely counterfactual explanations mirror the popularity profiles of the users who receive them.

\paragraph{Popularity Distribution Similarity (PDS)} Let $H$ be the complete collection of items that appear in users' interaction histories and $C$ the collection of items that appear in the generated counterfactual explanations. If an item occurs multiple times (e.g., across different users), it is included that many times in the collection. We convert each item to its global popularity score and place the scores into $B$ equal-width bins, producing two histograms (empirical distributions) $P_H$ and $P_C$. The similarity between the two popularity profiles can then be defined as the $\chi^2$ distance:
\begin{equation}
    \text{PDS} = \sum_{b=1}^{B} \frac{(P_H(b) - P_C(b))^2}{P_C(b) + \epsilon}
\end{equation}
\noindent with $\epsilon = 10^{-10}$ for numerical stability. A smaller PDS indicates that counterfactual items follow nearly the same popularity distribution as the historical items, suggesting lower popularity bias.

\paragraph{Expected Popularity Deviation (EPD)} EPD tracks the average shift in popularity at the user level. With $\bar{\pi}_{H_u}$ and $\bar{\pi}_{C_u}$ denoting the mean popularity of user $u$'s history and counterfactual set respectively:
\begin{equation}
    \text{EPD} = \frac{1}{|U|} \sum_{u \in U} (\bar{\pi}_{C_u} - \bar{\pi}_{H_u})^2
\end{equation}
Here we calculate the global EPD, but we can also use it locally for a single user by only computing $EPD_u = (\bar{\pi}_{C_u} - \bar{\pi}_{H_u})^2$. A small EPD indicates that a user receives counterfactual sets whose popularity closely matches what they historically prefer.

It should be noted that, as counterfactual sets are usually much smaller than the user history (often only one or two items), looking at the user level can yield volatile results. For this reason, we pool the users in PDS and examine the global EPD.

\subsection{Implementation details}
For our LLM-augmented CFEs, we employed Qwen3-8B, a recent open-weight language model. This 8 billion parameter instruction-tuned model, developed by Alibaba Group as part of the Qwen3 series, allows for both reasoning and traditional queries. In our experiments, we do not use the reasoning functionality. We use a quantized model (Q3\_K\_M) created by Unsloth~\cite{unsloth} to improve performance. For inference, we used the llama-cpp-python library. The model was not fine-tuned on multimedia data.

To compute the vector representations of the LLM-generated natural language profiles, we used Mixedbread's mxbai-embed-large-v1 model \cite{embed2024mxbai}. This open-weight model achieves state-of-the-art performance in its size class. All embedding operations were performed using the SentenceTransformers (SBERT)~\cite{reimers2019sentence}.
For recommendation, we used a standard Neural Matrix Factorization (NeuMF) model~\cite{ncf}. The model was trained with an embedding size of 16, a learning rate of 1e-3, and a weight decay of 1e-3. 

The implementation and hyperparameters of our recommendation model and counterfactual approach were based on the publicly available ACCENT codebase~\footnote{https://github.com/hieptk/accent}. We extended this baseline to LLM-augmented ACCENT. Our implementation can be found at \url{https://github.com/ahasami/llm-augmented-cfs}. 
\section{Results}

\begin{table}[t]
  \caption{Popularity alignment comparison of methods with respect to item groups on ML-1M and Amazon datasets.}
  \label{tab:results-alignment-split}
  \centering
  \small
  \setlength{\tabcolsep}{4pt}
  \begin{tabular}{l l rr rr}
    \toprule
        &         & \multicolumn{2}{c}{Niche} &
                      \multicolumn{2}{c}{Blockbuster} \\
    \cmidrule(lr){3-4}\cmidrule(lr){5-6}
    Dataset & Method &
    PDS~$\downarrow$ & EPD~$\downarrow$ &
    PDS~$\downarrow$ & EPD~$\downarrow$ \\
    \midrule
    \multirow{3}{*}{ML-1M}
      & Top Popular    & 1764.31 & 0.111 & 2588.44 & 0.070 \\
      & ACCENT & 8.58 & 0.012 & 21.58 & 0.017 \\
      & LLM-Augmented & \textbf{5.24} & \textbf{0.008} & \textbf{16.25} & \textbf{0.014} \\
    \midrule
    \multirow{3}{*}{Amazon}
      & Top Popular    & 158.88 & $5.1*10^{-7}$ & 165.87 & $4.2*10^{-5}$ \\
      & ACCENT & \textbf{3.83} & $2.0*10^{-8}$ & 14.29 & $1.4*10^{-5}$ \\
      & LLM-Augmented & 9.13 & \boldmath$1.9*10^{-8}$\unboldmath & \textbf{6.06} & \boldmath$1.4*10^{-5}$\unboldmath \\
    \bottomrule
  \end{tabular}
\end{table}


We begin by asking a simple question: \textit{does the LLM-augmentation step nudge counterfactual explanations towards the side of the popularity spectrum that each user actually favours?} To answer this, we generate ACCENT explanations with and without the filter. On ML-1M we allow the filter to drop $n=5$ history items per user, whereas on Amazon we cap it at $n=1$ so as not to wipe out the already thin behavioural signal. This dataset is a lot sparser than ML-1M, and a single deletion in a profile with few interactions can have a large impact. This sparsity also affects the performance of the recommendation model. We report nDCG@10 of 0.1303 and 0.0481 on ML-1M and Amazon respectively, computed using the negative sampling approach described in~\cite{ncf}. Here, for each test point of a user, 100 items the user has not interacted with are randomly sampled to form the candidate set. 



Table \ref{tab:results-alignment-split} shows that our method achieves the best scores across all groups, consistently lowering both PDS and EPD compared to ACCENT. However, most differences are not statistically significant, except the overall EPD on ML-1M (blockbuster and niche combined, not shown). Our approach outperforms ACCENT (0.017 vs 0.014, $p<0.05$). 

\begin{figure*}
    \centering
    \begin{subfigure}[b]{0.9\textwidth}
        \centering
        \includegraphics[width=\textwidth]{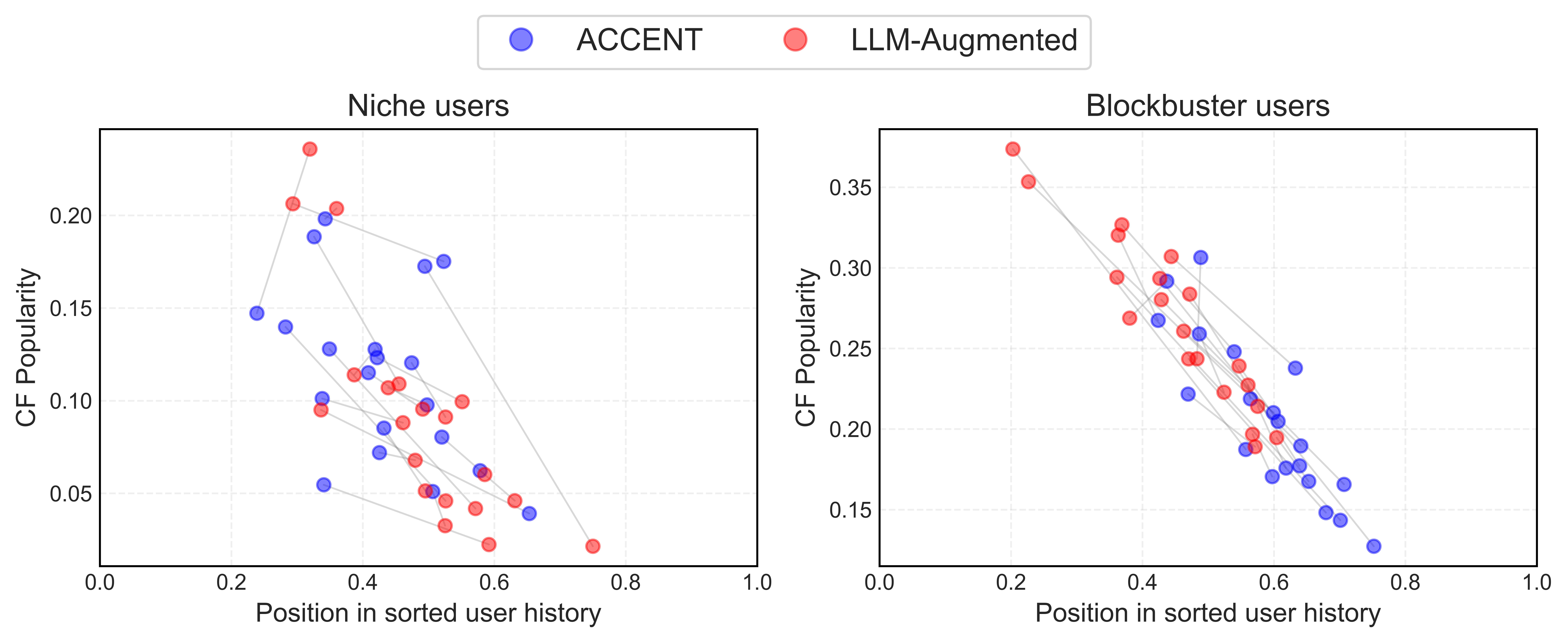}
        \caption{MovieLens 1M ($n=5$)}
        \label{figure:pop_vs_pos_ml}
    \end{subfigure}
    \hfill
    \begin{subfigure}[b]{0.9\textwidth}
        \centering
        \includegraphics[width=\textwidth]{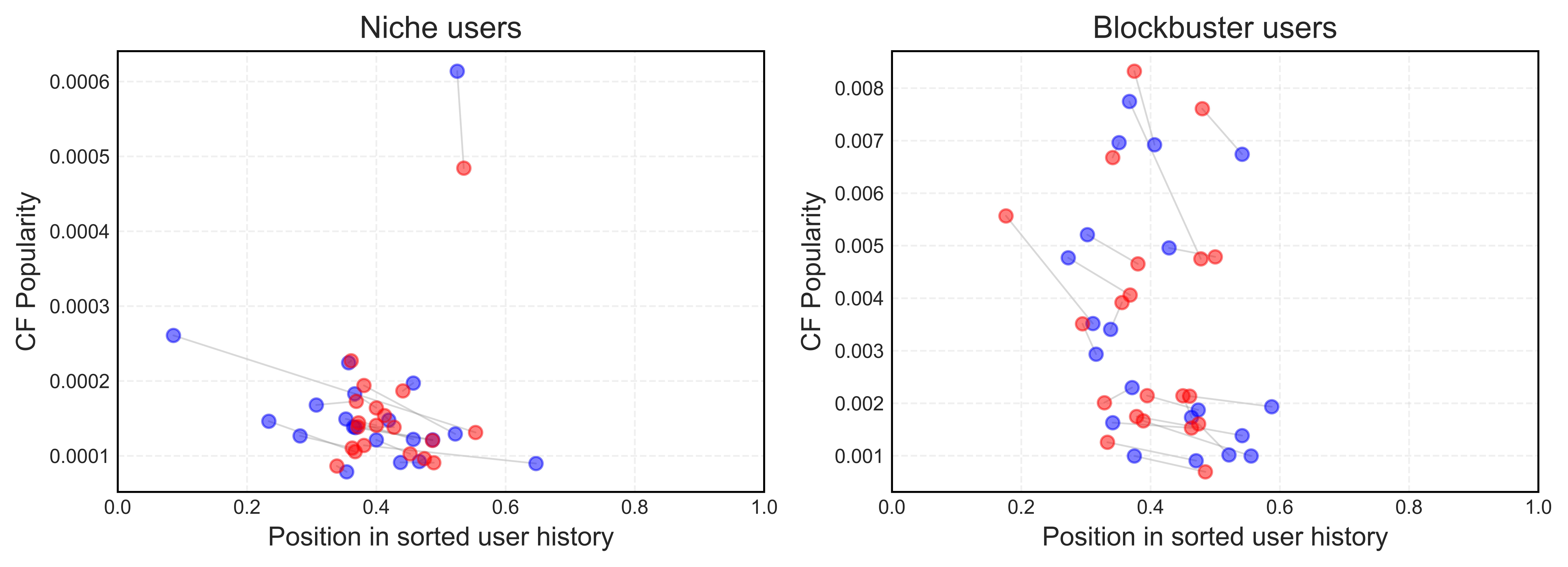}
        \caption{Amazon VideoGames ($n=1$)}
        \label{figure:pop_vs_pos_avg}
    \end{subfigure}
    \caption{Average popularity of counterfactual sets versus average position of counterfactual items in the normalized popularity-sorted user history. Each user category is binned into 20 bins based on the popularity of similar histories.}
    \label{fig:pop_vs_pos}
\end{figure*}

Fig. \ref{figure:pop_vs_pos_ml} shows a clear, consistent shift on ML-1M. For bins of niche users (left panel) our filtered counterfactuals move downward and slightly rightward, indicating less-popular items drawn from deeper in each user’s long-tail history. For the bins of blockbuster users (right panel) the same mechanism pushes explanations upward and leftward, highlighting even more popular items near the head of the profile. The result is a bidirectional correction. The same analysis on Amazon (Fig. \ref{figure:pop_vs_pos_avg}) is noisier. Some bins of blockbuster users benefit from a popularity boost, and for niche users the range of the user profile from which counterfactuals are generated is tightened.

We noticed a slight increase in cases where no CF could be found (ML-1M $14.56\% \rightarrow 21.16\%$, Amazon $27.09\% \rightarrow 27.31\%$). This makes sense, as we remove items that ACCENT can use to justify a change. By shrinking the candidate set, we sometimes eliminate the items that would have enabled a recommendation flip, especially for users whose profiles were minimal or highly concentrated.


Overall, the results show that LLM augmentation improves the popularity alignment of CFEs, especially on ML-1M, where user histories are richer and more stable. Bidirectional corrections are consistent: niche users get less popular explanations, while blockbuster users are nudged toward mainstream ones. Gains in EPD on ML-1M are small but statistically significant. On Amazon, impact is weaker—sparser histories limit the LLM’s ability to infer profiles and guide filtering. Still, the method never harms performance, suggesting it remains safe to apply even with limited effect.

\section{Discussion}
\label{section:discussion}
\paragraph{Limitations}
While our LLM-augmentation step shows promising improvements in user alignment, several limitations remain. In sparse datasets, the deletion of even a single item can disproportionately distort user profiles, and the LLM may struggle to infer coherent preferences. The LLM itself also introduces challenges: it can be inconsistent, generic or even hallucinatory in its summaries. Although our prompt design mitigates some of these issues, occasional failures still occur. The approach is also computationally expensive, as we need $n*h$ LLM queries per user, where $n$ is the number of iterations and $h$ the size of the user history. Finally, removing out-of-character items can increase the number of instances where ACCENT cannot find a valid CFE.

\paragraph{Future work}
Our method operates purely as a pre-processing step, without modifying the recommendation model or its outputs. Future work could look at how the recommendations would change, either by retraining or estimation through the usage of influence functions~\cite{bbpreds}. Also, exploring the interaction between LLM-augmented ACCENT and debiased recommendation outputs could offer insight into whether the observed biases in ACCENT stem from input data or the recommender itself. Finally, investigating the effects of larger models, or models fine-tuned on multimedia data, may further improve profile quality and filtering accuracy.
\section{Conclusion}
We introduced a straightforward, model-agnostic pre-processing step that uses large language models to improve the popularity alignment of counterfactual explanations in recommender systems. By filtering out history items inconsistent with a user's core preferences, we guide the explanation process toward outputs more representative of users. We specifically examined the effect of our method on popularity bias and showed that the approach improves popularity alignment, particularly in richer datasets like MovieLens. While benefits are less profound for sparser datasets, the method does not degrade performance and provides a promising foundation for further research into bias-aware explanation using LLMs.

\bibliographystyle{ACM-Reference-Format}
\bibliography{references}


\begin{thebibliography}{33}


\ifx \showCODEN    \undefined \def \showCODEN     #1{\unskip}     \fi
\ifx \showDOI      \undefined \def \showDOI       #1{#1}\fi
\ifx \showISBNx    \undefined \def \showISBNx     #1{\unskip}     \fi
\ifx \showISBNxiii \undefined \def \showISBNxiii  #1{\unskip}     \fi
\ifx \showISSN     \undefined \def \showISSN      #1{\unskip}     \fi
\ifx \showLCCN     \undefined \def \showLCCN      #1{\unskip}     \fi
\ifx \shownote     \undefined \def \shownote      #1{#1}          \fi
\ifx \showarticletitle \undefined \def \showarticletitle #1{#1}   \fi
\ifx \showURL      \undefined \def \showURL       {\relax}        \fi
\providecommand\bibfield[2]{#2}
\providecommand\bibinfo[2]{#2}
\providecommand\natexlab[1]{#1}
\providecommand\showeprint[2][]{arXiv:#2}

\bibitem[Abdollahpouri et~al\mbox{.}(2019)]%
        {unfairness_of_pop_bias}
\bibfield{author}{\bibinfo{person}{Himan Abdollahpouri}, \bibinfo{person}{Masoud Mansoury}, \bibinfo{person}{Robin Burke}, {and} \bibinfo{person}{Bamshad Mobasher}.} \bibinfo{year}{2019}\natexlab{}.
\newblock \bibinfo{title}{The Unfairness of Popularity Bias in Recommendation}.
\newblock
\newblock
\showeprint[arxiv]{1907.13286}~[cs.IR]
\urldef\tempurl%
\url{https://arxiv.org/abs/1907.13286}
\showURL{%
\tempurl}


\bibitem[Atzmueller et~al\mbox{.}(2024)]%
        {expl_machine_learning}
\bibfield{author}{\bibinfo{person}{Martin Atzmueller}, \bibinfo{person}{Johannes F{\"u}rnkranz}, \bibinfo{person}{Tom{\'a}{\v{s}} Kliegr}, {and} \bibinfo{person}{Ute Schmid}.} \bibinfo{year}{2024}\natexlab{}.
\newblock \showarticletitle{Explainable and interpretable machine learning and data mining}.
\newblock \bibinfo{journal}{\emph{Data Mining and Knowledge Discovery}} \bibinfo{volume}{38}, \bibinfo{number}{5} (\bibinfo{date}{01 Sep} \bibinfo{year}{2024}), \bibinfo{pages}{2571--2595}.
\newblock
\showISSN{1573-756X}
\urldef\tempurl%
\url{https://doi.org/10.1007/s10618-024-01041-y}
\showDOI{\tempurl}


\bibitem[Chen et~al\mbox{.}(2022)]%
        {grease}
\bibfield{author}{\bibinfo{person}{Ziheng Chen}, \bibinfo{person}{Fabrizio Silvestri}, \bibinfo{person}{Jia Wang}, \bibinfo{person}{Yongfeng Zhang}, \bibinfo{person}{Zhenhua Huang}, \bibinfo{person}{Hongshik Ahn}, {and} \bibinfo{person}{Gabriele Tolomei}.} \bibinfo{year}{2022}\natexlab{}.
\newblock \bibinfo{title}{GREASE: Generate Factual and Counterfactual Explanations for GNN-based Recommendations}.
\newblock
\newblock
\showeprint[arxiv]{2208.04222}~[cs.IR]
\urldef\tempurl%
\url{https://arxiv.org/abs/2208.04222}
\showURL{%
\tempurl}


\bibitem[Cheng et~al\mbox{.}(2019)]%
        {fastinfanal}
\bibfield{author}{\bibinfo{person}{Weiyu Cheng}, \bibinfo{person}{Yanyan Shen}, \bibinfo{person}{Linpeng Huang}, {and} \bibinfo{person}{Yanmin Zhu}.} \bibinfo{year}{2019}\natexlab{}.
\newblock \showarticletitle{Incorporating Interpretability into Latent Factor Models via Fast Influence Analysis}. In \bibinfo{booktitle}{\emph{Proceedings of the 25th ACM SIGKDD International Conference on Knowledge Discovery \& Data Mining}} (Anchorage, AK, USA) \emph{(\bibinfo{series}{KDD '19})}. \bibinfo{publisher}{Association for Computing Machinery}, \bibinfo{address}{New York, NY, USA}, \bibinfo{pages}{885–893}.
\newblock
\showISBNx{9781450362016}
\urldef\tempurl%
\url{https://doi.org/10.1145/3292500.3330857}
\showDOI{\tempurl}


\bibitem[Cramer et~al\mbox{.}(2008)]%
        {effect_transparancy_on_trust_art}
\bibfield{author}{\bibinfo{person}{Henriette Cramer}, \bibinfo{person}{Vanessa Evers}, \bibinfo{person}{Satyan Ramlal}, \bibinfo{person}{Maarten van Someren}, \bibinfo{person}{Lloyd Rutledge}, \bibinfo{person}{Natalia Stash}, \bibinfo{person}{Lora Aroyo}, {and} \bibinfo{person}{Bob Wielinga}.} \bibinfo{year}{2008}\natexlab{}.
\newblock \showarticletitle{The effects of transparency on trust in and acceptance of a content-based art recommender}.
\newblock \bibinfo{journal}{\emph{User Modeling and User-Adapted Interaction}} \bibinfo{volume}{18}, \bibinfo{number}{5} (\bibinfo{date}{01 Nov} \bibinfo{year}{2008}), \bibinfo{pages}{455--496}.
\newblock
\showISSN{1573-1391}
\urldef\tempurl%
\url{https://doi.org/10.1007/s11257-008-9051-3}
\showDOI{\tempurl}


\bibitem[Daniel~Han and team(2023)]%
        {unsloth}
\bibfield{author}{\bibinfo{person}{Michael~Han Daniel~Han} {and} \bibinfo{person}{Unsloth team}.} \bibinfo{year}{2023}\natexlab{}.
\newblock \bibinfo{booktitle}{\emph{Unsloth}}.
\newblock
\urldef\tempurl%
\url{http://github.com/unslothai/unsloth}
\showURL{%
\tempurl}


\bibitem[Deldjoo et~al\mbox{.}(2020)]%
        {recsys_multimedia}
\bibfield{author}{\bibinfo{person}{Yashar Deldjoo}, \bibinfo{person}{Markus Schedl}, \bibinfo{person}{Paolo Cremonesi}, {and} \bibinfo{person}{Gabriella Pasi}.} \bibinfo{year}{2020}\natexlab{}.
\newblock \showarticletitle{Recommender Systems Leveraging Multimedia Content}.
\newblock \bibinfo{journal}{\emph{Comput. Surveys}} \bibinfo{volume}{53}, \bibinfo{number}{5}, Article \bibinfo{articleno}{106} (\bibinfo{date}{Sept.} \bibinfo{year}{2020}), \bibinfo{numpages}{38}~pages.
\newblock
\showISSN{0360-0300}
\urldef\tempurl%
\url{https://doi.org/10.1145/3407190}
\showDOI{\tempurl}


\bibitem[Dong et~al\mbox{.}(2022)]%
        {trust_deep_learning}
\bibfield{author}{\bibinfo{person}{Manqing Dong}, \bibinfo{person}{Feng Yuan}, \bibinfo{person}{Lina Yao}, \bibinfo{person}{Xianzhi Wang}, \bibinfo{person}{Xiwei Xu}, {and} \bibinfo{person}{Liming Zhu}.} \bibinfo{year}{2022}\natexlab{}.
\newblock \showarticletitle{A survey for trust-aware recommender systems: A deep learning perspective}.
\newblock \bibinfo{journal}{\emph{Knowledge-Based Systems}}  \bibinfo{volume}{249} (\bibinfo{year}{2022}), \bibinfo{pages}{108954}.
\newblock
\showISSN{0950-7051}
\urldef\tempurl%
\url{https://doi.org/10.1016/j.knosys.2022.108954}
\showDOI{\tempurl}


\bibitem[Ge et~al\mbox{.}(2024)]%
        {xrec_survey}
\bibfield{author}{\bibinfo{person}{Yingqiang Ge}, \bibinfo{person}{Shuchang Liu}, \bibinfo{person}{Zuohui Fu}, \bibinfo{person}{Juntao Tan}, \bibinfo{person}{Zelong Li}, \bibinfo{person}{Shuyuan Xu}, \bibinfo{person}{Yunqi Li}, \bibinfo{person}{Yikun Xian}, {and} \bibinfo{person}{Yongfeng Zhang}.} \bibinfo{year}{2024}\natexlab{}.
\newblock \showarticletitle{A Survey on Trustworthy Recommender Systems}.
\newblock \bibinfo{journal}{\emph{ACM Trans. Recomm. Syst.}} \bibinfo{volume}{3}, \bibinfo{number}{2}, Article \bibinfo{articleno}{13} (\bibinfo{date}{Nov.} \bibinfo{year}{2024}), \bibinfo{numpages}{68}~pages.
\newblock
\urldef\tempurl%
\url{https://doi.org/10.1145/3652891}
\showDOI{\tempurl}


\bibitem[Ghazimatin et~al\mbox{.}(2020)]%
        {prince}
\bibfield{author}{\bibinfo{person}{Azin Ghazimatin}, \bibinfo{person}{Oana Balalau}, \bibinfo{person}{Rishiraj Saha~Roy}, {and} \bibinfo{person}{Gerhard Weikum}.} \bibinfo{year}{2020}\natexlab{}.
\newblock \showarticletitle{PRINCE: Provider-side Interpretability with Counterfactual Explanations in Recommender Systems}. In \bibinfo{booktitle}{\emph{Proceedings of the 13th International Conference on Web Search and Data Mining}} (Houston, TX, USA) \emph{(\bibinfo{series}{WSDM '20})}. \bibinfo{publisher}{Association for Computing Machinery}, \bibinfo{address}{New York, NY, USA}, \bibinfo{pages}{196–204}.
\newblock
\showISBNx{9781450368223}
\urldef\tempurl%
\url{https://doi.org/10.1145/3336191.3371824}
\showDOI{\tempurl}


\bibitem[Hampel(1974)]%
        {influence_curve}
\bibfield{author}{\bibinfo{person}{Frank~R. Hampel}.} \bibinfo{year}{1974}\natexlab{}.
\newblock \showarticletitle{The Influence Curve and its Role in Robust Estimation}.
\newblock \bibinfo{journal}{\emph{J. Amer. Statist. Assoc.}} \bibinfo{volume}{69}, \bibinfo{number}{346} (\bibinfo{year}{1974}), \bibinfo{pages}{383--393}.
\newblock
\urldef\tempurl%
\url{https://doi.org/10.1080/01621459.1974.10482962}
\showDOI{\tempurl}
\showeprint{https://www.tandfonline.com/doi/pdf/10.1080/01621459.1974.10482962}


\bibitem[Harper and Konstan(2015)]%
        {movielens}
\bibfield{author}{\bibinfo{person}{F.~Maxwell Harper} {and} \bibinfo{person}{Joseph~A. Konstan}.} \bibinfo{year}{2015}\natexlab{}.
\newblock \showarticletitle{The MovieLens Datasets: History and Context}.
\newblock \bibinfo{journal}{\emph{ACM Trans. Interact. Intell. Syst.}} \bibinfo{volume}{5}, \bibinfo{number}{4}, Article \bibinfo{articleno}{19} (\bibinfo{date}{Dec.} \bibinfo{year}{2015}), \bibinfo{numpages}{19}~pages.
\newblock
\showISSN{2160-6455}
\urldef\tempurl%
\url{https://doi.org/10.1145/2827872}
\showDOI{\tempurl}


\bibitem[He et~al\mbox{.}(2017)]%
        {ncf}
\bibfield{author}{\bibinfo{person}{Xiangnan He}, \bibinfo{person}{Lizi Liao}, \bibinfo{person}{Hanwang Zhang}, \bibinfo{person}{Liqiang Nie}, \bibinfo{person}{Xia Hu}, {and} \bibinfo{person}{Tat-Seng Chua}.} \bibinfo{year}{2017}\natexlab{}.
\newblock \showarticletitle{Neural Collaborative Filtering}. In \bibinfo{booktitle}{\emph{Proceedings of the 26th International Conference on World Wide Web}} (Perth, Australia) \emph{(\bibinfo{series}{WWW '17})}. \bibinfo{publisher}{International World Wide Web Conferences Steering Committee}, \bibinfo{address}{Republic and Canton of Geneva, CHE}, \bibinfo{pages}{173–182}.
\newblock
\showISBNx{9781450349130}
\urldef\tempurl%
\url{https://doi.org/10.1145/3038912.3052569}
\showDOI{\tempurl}


\bibitem[Hou et~al\mbox{.}(2024)]%
        {hou2024bridging}
\bibfield{author}{\bibinfo{person}{Yupeng Hou}, \bibinfo{person}{Jiacheng Li}, \bibinfo{person}{Zhankui He}, \bibinfo{person}{An Yan}, \bibinfo{person}{Xiusi Chen}, {and} \bibinfo{person}{Julian McAuley}.} \bibinfo{year}{2024}\natexlab{}.
\newblock \bibinfo{title}{Bridging Language and Items for Retrieval and Recommendation}.
\newblock
\newblock
\showeprint[arxiv]{2403.03952}~[cs.IR]
\urldef\tempurl%
\url{https://arxiv.org/abs/2403.03952}
\showURL{%
\tempurl}


\bibitem[Klimashevskaia et~al\mbox{.}(2024)]%
        {klimashevskaia2024survey}
\bibfield{author}{\bibinfo{person}{Anastasiia Klimashevskaia}, \bibinfo{person}{Dietmar Jannach}, \bibinfo{person}{Mehdi Elahi}, {and} \bibinfo{person}{Christoph Trattner}.} \bibinfo{year}{2024}\natexlab{}.
\newblock \showarticletitle{A survey on popularity bias in recommender systems}.
\newblock \bibinfo{journal}{\emph{User Modeling and User-Adapted Interaction}} \bibinfo{volume}{34}, \bibinfo{number}{5} (\bibinfo{year}{2024}), \bibinfo{pages}{1777--1834}.
\newblock


\bibitem[Koh and Liang(2017)]%
        {bbpreds}
\bibfield{author}{\bibinfo{person}{Pang~Wei Koh} {and} \bibinfo{person}{Percy Liang}.} \bibinfo{year}{2017}\natexlab{}.
\newblock \showarticletitle{Understanding black-box predictions via influence functions}. In \bibinfo{booktitle}{\emph{Proceedings of the 34th International Conference on Machine Learning - Volume 70}} (Sydney, NSW, Australia) \emph{(\bibinfo{series}{ICML'17})}. \bibinfo{publisher}{JMLR.org}, \bibinfo{pages}{1885–1894}.
\newblock


\bibitem[Lee et~al\mbox{.}(2024)]%
        {embed2024mxbai}
\bibfield{author}{\bibinfo{person}{Sean Lee}, \bibinfo{person}{Aamir Shakir}, \bibinfo{person}{Darius Koenig}, {and} \bibinfo{person}{Julius Lipp}.} \bibinfo{year}{2024}\natexlab{}.
\newblock \bibinfo{booktitle}{\emph{Open Source Strikes Bread - New Fluffy Embedding Model}}.
\newblock
\urldef\tempurl%
\url{https://www.mixedbread.ai/blog/mxbai-embed-large-v1}
\showURL{%
\tempurl}


\bibitem[Lichtenberg et~al\mbox{.}(2024)]%
        {llms_pop_bias}
\bibfield{author}{\bibinfo{person}{Jan~Malte Lichtenberg}, \bibinfo{person}{Alexander Buchholz}, {and} \bibinfo{person}{Pola Schwöbel}.} \bibinfo{year}{2024}\natexlab{}.
\newblock \bibinfo{title}{Large language models as recommender systems: A study of popularity bias}.
\newblock
\newblock
\urldef\tempurl%
\url{https://www.amazon.science/publications/large-language-models-as-recommender-systems-a-study-of-popularity-bias}
\showURL{%
\tempurl}


\bibitem[Lucic et~al\mbox{.}(2022)]%
        {cf-gnn-explainer}
\bibfield{author}{\bibinfo{person}{Ana Lucic}, \bibinfo{person}{Maartje ter Hoeve}, \bibinfo{person}{Gabriele Tolomei}, \bibinfo{person}{Maarten de Rijke}, {and} \bibinfo{person}{Fabrizio Silvestri}.} \bibinfo{year}{2022}\natexlab{}.
\newblock \bibinfo{title}{CF-GNNExplainer: Counterfactual Explanations for Graph Neural Networks}.
\newblock
\newblock
\showeprint[arxiv]{2102.03322}~[cs.LG]
\urldef\tempurl%
\url{https://arxiv.org/abs/2102.03322}
\showURL{%
\tempurl}


\bibitem[Mansoury(2022)]%
        {mansoury2022understanding}
\bibfield{author}{\bibinfo{person}{Masoud Mansoury}.} \bibinfo{year}{2022}\natexlab{}.
\newblock \showarticletitle{Understanding and mitigating multi-sided exposure bias in recommender systems}.
\newblock \bibinfo{journal}{\emph{ACM SIGWEB Newsletter}} \bibinfo{volume}{2022}, \bibinfo{number}{Autumn} (\bibinfo{year}{2022}), \bibinfo{pages}{1--4}.
\newblock


\bibitem[Mansoury et~al\mbox{.}(2020)]%
        {mansoury2020feedback}
\bibfield{author}{\bibinfo{person}{Masoud Mansoury}, \bibinfo{person}{Himan Abdollahpouri}, \bibinfo{person}{Mykola Pechenizkiy}, \bibinfo{person}{Bamshad Mobasher}, {and} \bibinfo{person}{Robin Burke}.} \bibinfo{year}{2020}\natexlab{}.
\newblock \showarticletitle{Feedback loop and bias amplification in recommender systems}. In \bibinfo{booktitle}{\emph{Proceedings of the 29th ACM international conference on information \& knowledge management}}. \bibinfo{pages}{2145--2148}.
\newblock


\bibitem[Medda et~al\mbox{.}(2024)]%
        {GNNUERS}
\bibfield{author}{\bibinfo{person}{Giacomo Medda}, \bibinfo{person}{Francesco Fabbri}, \bibinfo{person}{Mirko Marras}, \bibinfo{person}{Ludovico Boratto}, {and} \bibinfo{person}{Gianni Fenu}.} \bibinfo{year}{2024}\natexlab{}.
\newblock \showarticletitle{GNNUERS: Fairness Explanation in GNNs for Recommendation via Counterfactual Reasoning}.
\newblock \bibinfo{journal}{\emph{ACM Trans. Intell. Syst. Technol.}} \bibinfo{volume}{16}, \bibinfo{number}{1}, Article \bibinfo{articleno}{6} (\bibinfo{date}{Dec.} \bibinfo{year}{2024}), \bibinfo{numpages}{26}~pages.
\newblock
\showISSN{2157-6904}
\urldef\tempurl%
\url{https://doi.org/10.1145/3655631}
\showDOI{\tempurl}


\bibitem[Molnar(2025)]%
        {interpretable_machine_learning}
\bibfield{author}{\bibinfo{person}{Christoph Molnar}.} \bibinfo{year}{2025}\natexlab{}.
\newblock \bibinfo{booktitle}{\emph{Interpretable Machine Learning} (\bibinfo{edition}{3} ed.)}.
\newblock \bibinfo{publisher}{Independently published}.
\newblock
\showISBNx{978-3-911578-03-5}
\urldef\tempurl%
\url{https://christophm.github.io/interpretable-ml-book}
\showURL{%
\tempurl}


\bibitem[Mothilal et~al\mbox{.}(2019)]%
        {DiCE}
\bibfield{author}{\bibinfo{person}{Ramaravind~Kommiya Mothilal}, \bibinfo{person}{Amit Sharma}, {and} \bibinfo{person}{Chenhao Tan}.} \bibinfo{year}{2019}\natexlab{}.
\newblock \showarticletitle{Explaining Machine Learning Classifiers through Diverse Counterfactual Explanations}.
\newblock \bibinfo{journal}{\emph{CoRR}}  \bibinfo{volume}{abs/1905.07697} (\bibinfo{year}{2019}).
\newblock
\showeprint[arXiv]{1905.07697}
\urldef\tempurl%
\url{http://arxiv.org/abs/1905.07697}
\showURL{%
\tempurl}


\bibitem[Prent and Mansoury(2024)]%
        {prent2024correcting}
\bibfield{author}{\bibinfo{person}{Juno Prent} {and} \bibinfo{person}{Masoud Mansoury}.} \bibinfo{year}{2024}\natexlab{}.
\newblock \showarticletitle{Correcting Popularity Bias in Recommender Systems via Item Loss Equalization}.
\newblock \bibinfo{journal}{\emph{arXiv preprint arXiv:2410.04830}} (\bibinfo{year}{2024}).
\newblock


\bibitem[Reimers and Gurevych(2019)]%
        {reimers2019sentence}
\bibfield{author}{\bibinfo{person}{Nils Reimers} {and} \bibinfo{person}{Iryna Gurevych}.} \bibinfo{year}{2019}\natexlab{}.
\newblock \showarticletitle{Sentence-bert: Sentence embeddings using siamese bert-networks}.
\newblock \bibinfo{journal}{\emph{arXiv preprint arXiv:1908.10084}} (\bibinfo{year}{2019}).
\newblock


\bibitem[Sabouri et~al\mbox{.}(2025)]%
        {temporal_user_profiling}
\bibfield{author}{\bibinfo{person}{Milad Sabouri}, \bibinfo{person}{Masoud Mansoury}, \bibinfo{person}{Kun Lin}, {and} \bibinfo{person}{Bamshad Mobasher}.} \bibinfo{year}{2025}\natexlab{}.
\newblock \showarticletitle{Towards Explainable Temporal User Profiling with LLMs}. In \bibinfo{booktitle}{\emph{Adjunct Proceedings of the 33rd ACM Conference on User Modeling, Adaptation and Personalization}} \emph{(\bibinfo{series}{UMAP Adjunct '25})}. \bibinfo{publisher}{Association for Computing Machinery}, \bibinfo{address}{New York, NY, USA}, \bibinfo{pages}{219–227}.
\newblock
\showISBNx{9798400713996}
\urldef\tempurl%
\url{https://doi.org/10.1145/3708319.3733655}
\showDOI{\tempurl}


\bibitem[Schafer et~al\mbox{.}(1999)]%
        {recsys_ecommerce}
\bibfield{author}{\bibinfo{person}{J.~Ben Schafer}, \bibinfo{person}{Joseph Konstan}, {and} \bibinfo{person}{John Riedl}.} \bibinfo{year}{1999}\natexlab{}.
\newblock \showarticletitle{Recommender systems in e-commerce}. In \bibinfo{booktitle}{\emph{Proceedings of the 1st ACM Conference on Electronic Commerce}} (Denver, Colorado, USA) \emph{(\bibinfo{series}{EC '99})}. \bibinfo{publisher}{Association for Computing Machinery}, \bibinfo{address}{New York, NY, USA}, \bibinfo{pages}{158–166}.
\newblock
\showISBNx{1581131763}
\urldef\tempurl%
\url{https://doi.org/10.1145/336992.337035}
\showDOI{\tempurl}


\bibitem[Shang et~al\mbox{.}(2022)]%
        {why_am_i_not_seeing_it}
\bibfield{author}{\bibinfo{person}{Ruoxi Shang}, \bibinfo{person}{K.~J.~Kevin Feng}, {and} \bibinfo{person}{Chirag Shah}.} \bibinfo{year}{2022}\natexlab{}.
\newblock \showarticletitle{Why Am I Not Seeing It? Understanding Users’ Needs for Counterfactual Explanations in Everyday Recommendations}. In \bibinfo{booktitle}{\emph{Proceedings of the 2022 ACM Conference on Fairness, Accountability, and Transparency}} (Seoul, Republic of Korea) \emph{(\bibinfo{series}{FAccT '22})}. \bibinfo{publisher}{Association for Computing Machinery}, \bibinfo{address}{New York, NY, USA}, \bibinfo{pages}{1330–1340}.
\newblock
\showISBNx{9781450393522}
\urldef\tempurl%
\url{https://doi.org/10.1145/3531146.3533189}
\showDOI{\tempurl}


\bibitem[Tan et~al\mbox{.}(2021)]%
        {countER}
\bibfield{author}{\bibinfo{person}{Juntao Tan}, \bibinfo{person}{Shuyuan Xu}, \bibinfo{person}{Yingqiang Ge}, \bibinfo{person}{Yunqi Li}, \bibinfo{person}{Xu Chen}, {and} \bibinfo{person}{Yongfeng Zhang}.} \bibinfo{year}{2021}\natexlab{}.
\newblock \showarticletitle{Counterfactual Explainable Recommendation}. In \bibinfo{booktitle}{\emph{Proceedings of the 30th ACM International Conference on Information \& Knowledge Management}} (Virtual Event, Queensland, Australia) \emph{(\bibinfo{series}{CIKM '21})}. \bibinfo{publisher}{Association for Computing Machinery}, \bibinfo{address}{New York, NY, USA}, \bibinfo{pages}{1784–1793}.
\newblock
\showISBNx{9781450384469}
\urldef\tempurl%
\url{https://doi.org/10.1145/3459637.3482420}
\showDOI{\tempurl}


\bibitem[Tintarev and Masthoff(2007)]%
        {effective_explanations}
\bibfield{author}{\bibinfo{person}{Nava Tintarev} {and} \bibinfo{person}{Judith Masthoff}.} \bibinfo{year}{2007}\natexlab{}.
\newblock \showarticletitle{Effective explanations of recommendations: user-centered design}. In \bibinfo{booktitle}{\emph{Proceedings of the 2007 ACM Conference on Recommender Systems}} (Minneapolis, MN, USA) \emph{(\bibinfo{series}{RecSys '07})}. \bibinfo{publisher}{Association for Computing Machinery}, \bibinfo{address}{New York, NY, USA}, \bibinfo{pages}{153–156}.
\newblock
\showISBNx{9781595937308}
\urldef\tempurl%
\url{https://doi.org/10.1145/1297231.1297259}
\showDOI{\tempurl}


\bibitem[Tran et~al\mbox{.}(2021)]%
        {accent}
\bibfield{author}{\bibinfo{person}{Khanh~Hiep Tran}, \bibinfo{person}{Azin Ghazimatin}, {and} \bibinfo{person}{Rishiraj Saha~Roy}.} \bibinfo{year}{2021}\natexlab{}.
\newblock \showarticletitle{Counterfactual Explanations for Neural Recommenders}. In \bibinfo{booktitle}{\emph{Proceedings of the 44th International ACM SIGIR Conference on Research and Development in Information Retrieval}} (Virtual Event, Canada) \emph{(\bibinfo{series}{SIGIR '21})}. \bibinfo{publisher}{Association for Computing Machinery}, \bibinfo{address}{New York, NY, USA}, \bibinfo{pages}{1627–1631}.
\newblock
\showISBNx{9781450380379}
\urldef\tempurl%
\url{https://doi.org/10.1145/3404835.3463005}
\showDOI{\tempurl}


\bibitem[Wachter et~al\mbox{.}(2018)]%
        {cf_exs}
\bibfield{author}{\bibinfo{person}{Sandra Wachter}, \bibinfo{person}{Brent Mittelstadt}, {and} \bibinfo{person}{Chris Russell}.} \bibinfo{year}{2018}\natexlab{}.
\newblock \bibinfo{title}{Counterfactual Explanations without Opening the Black Box: Automated Decisions and the GDPR}.
\newblock
\newblock
\showeprint[arxiv]{1711.00399}~[cs.AI]
\urldef\tempurl%
\url{https://arxiv.org/abs/1711.00399}
\showURL{%
\tempurl}


\end{thebibliography}
\end{document}